# Anisotropic Lattice Expansion of Monolayer $WSe_2$ Revealed by Ultrafast Electron Diffraction

*Tony E. Karam[1,2] and Geoffrey A. Blake[1,3]\**

[1]Division of Chemistry and Chemical Engineering, California Institute of Technology Pasadena, CA 91125, USA; [2]IGP Photonics, Marlborough, MA 01752, USA  [3]Division of Geological and Planetary Sciences, California Institute of Technology, Pasadena, CA 91125, USA

**Abstract.** Bulk layered $MX_2$ transition metal chalcogenides (M = Mo, W and X = S, Se) are known to exhibit an indirect to direct bandgap transition as the number of layers is reduced. Previous time-resolved work has principally focused on the investigation of the transient evolution of the band structure after photo-excitation, but additional information on the dynamics of the concomitant lattice rearrangement is needed to fully understand this phenomenon. Here, ultrafast electron diffraction is used to probe the atomic motion and bond dilation in monolayer $WSe_2$ with femtosecond temporal resolution. The change in the intensity of the Bragg diffraction spots is characterized by single-exponential dynamics, consistent with a collective response of the lattice during electron-phonon and phonon-phonon equilibration that is repeatable over many hours of illumination with ultrafast pulses. Moreover, a transient and highly anisotropic lattice expansion is observed. A possible explanation for this behavior is axial strain induced by the laser excitation that breaks the degeneracy of the in-plane phonon modes, and that strongly influences the electronic band structure. Such degeneracy-breaking induced by distortions in the lattice are well characterized in static strain measurements, and the results presented here provide valuable insights into the nature and time scales of the structural rearrangement that occurs following optical photo-excitation in monolayer tungsten diselenide.

# Introduction

Two-dimensional (2D) materials have been widely investigated over the last decade since the first exfoliation of graphene, thanks in large measure to their numerous potential applications that span optoelectronics to biosensing. (1,2) In particular, 2D transition metal dichalcogenides (TMDCs) possess bandgaps on the order of 1.6-2.1 eV, leading to unique mechanical, chemical, electronic and thermal properties. (3-7) A direct to indirect bandgap shift is observed in thinned $MX_2$ materials (M = Mo, W and X = S, Se), and arises from shifts of the valence band hills and conduction band valleys in the Brillouin zone. (8-11) Accordingly, the generation and quantum manipulation of valley coherences in TMDC monolayers has been extensively studied for applications in valleytronics. (12-17) In particular, the confinement of electrons and holes to the $\pm K$ valleys and the valley degrees of freedom in these materials creates an unusual band structure (12) that leads to tunable and unique optical properties. (18-24)

First principles calculations have suggested that edge dislocation in single layer transition metal disulfides $MS_2$ (M=Mo or W) leads to local distortions of the crystal lattice, for example. (25) This in turn can have remarkable effects on the photoluminescence and electrical conductivity (26), and can be used to probe the band structure of the monolayer. Despite the remarkable progress achieved experimentally and theoretically on the electronic properties of TMDCs, the concomitant ultrafast structural dynamics resulting from the unique band structure of these materials remain poorly characterized. Such lattice dynamics are essential for the manipulation and control of the optical, electronic, and mechanical properties of these monolayers. While Raman spectroscopy has been previously used to study the temperature-dependent anharmonic lattice expansion of atomically-thin TMDCs semiconductors, such an approach yields only limited information on nuclear dynamics.

Ultrafast electron diffraction (UED) (27-34) provides a unique tool for the study of transient structural dynamics in monolayers and atomically-thin samples due to the large scattering cross section of electrons. In this technique, an optical pump pulse initiates the dynamics in the sample, and is followed by sub-picosecond electron pulses that probe the structural changes at various pump-probe time delays. Using UED, processes such as correlated atomic motions, bond dilation, and structural transformation in nanostructures can be investigated at a spatial and temporal resolution commensurate with the nuclear motion. (35-39)

In this manuscript, the ultrafast structural dynamics of WSe$_2$ monolayer are investigated using UED. The suppression of the intensity of the Bragg diffraction spots follow a single-exponential process that can be modeled using a single-mode relaxation time approximation. A fluence-dependent anisotropic expansion of the crystal lattice resulting from in-plane anharmonic phonon modes is observed, following excitation in resonance with the photoluminescence band of monolayer WSe$_2$. These results provide valuable insights into the time scales and symmetry of the ultrafast lattice rearrangement that occurs during the indirect optical transition induced by the evolution of the band structure. Ultimately, the control of these lattice dynamics may potentially tailor the optoelectronic properties of 2D semiconductor TMDCs on ultrafast time scales.

## Methods

### Sample Characterization and Preparation

The CVD-grown WSe$_2$ monolayers studied here were obtained from 2DLayer and transferred to TEM grids. To characterize the sample, photoluminescence and Raman spectra of representative flakes were acquired, and are displayed Figure 1. The photoluminescence spectrum is obtained with 532 nm, 0.1 mW laser excitation. A strong emission band centered at 790 nm that extends from 730 - 860 nm is detected, and is indicative of the indirect to direct bandgap transition in the monolayer. (8,10,40) The emission peak is very slightly red shifted from an unstrained monolayer sample, and indicates an axial strain of <0.5% for samples mounted on the TEM grids. The photoluminescence band originates from a K → Γ transition, where the conduction band minimum is located at the K point. (41) With excitation at 514 nm, we observe an intense (unpolarized) Raman peak centered at 250 cm$^{-1}$, followed by a smaller shoulder centered at 260 cm$^{-1}$. These are attributed to the $E_{2g}^1$ and $2LA$ phonon modes, respectively. (42)

Figure 2a shows a schematic of the UED setup, in which ultrafast laser pump and electron probe beams are overlapped onto a TEM grid containing the sample in transmission geometry. An atomic force microscopy (AFM) image of the WSe$_2$ monolayer flake deposited on sapphire is shown in Figure 2b, from which a flake thickness of ~0.65 nm is measured, corresponding to the height of a single unit cell (Figure 2c). The WSe$_2$ samples deposited on the TEM grids are annealed in a tube furnace at 160 °C for 2 hours at a pressure of 12 mTorr immediately before they were loaded into the ultra-high vacuum chamber for analysis.

**Ultrafast Electron Diffraction Apparatus**

The UED setup (see Fig. 2), described previously, (43-45) consists of a Ti:Sapphire oscillator and regenerative amplifier that generates 800 nm pulses of 100 fs duration at a repetition rate of 2 kHz. The fundamental beam is split into pump and probe arms using an ultrafast beam splitter. The pump pulses are P-polarized using a polarizing beam cube and are focused onto the sample that is housed in a UHV chamber. The probe beam pass through a frequency tripler to generate 266 nm UV pulses that are subsequently focused onto a $LaB_6$ photocathode. The generated photoelectrons are then accelerated to 30 keV and tightly focused to a spatial overlap with pump beam at the sample surface. To minimize space-charge effects, each electron pulse contains ~300 photoelectrons. The spatial overlap of the pump/probe beams is ensured by maximizing the transmission of both beams through a 150 μm dimeter circular aperture at the sample plane. The sample is mounted on a computer-controlled five-axis stage, and the TEM grid is aligned to be perpendicular to the electron beam axis to within a fraction of a degree. The pump-probe sweep is generated using a delay line, and the time-dependent diffraction patterns are recorded using a microchannel plate/phosphor screen coupled to a CCD working in the gate mode. To guard against unintended transient (heating) effects, the flake was exposed to the full laser pump fluence for several hours before beginning the UED measurements.

**Results**

Here the transient structural dynamics of monolayer $WSe_2$ are studied with UED driven by 800 nm pump pulses, which are resonant with the indirect bandgap (c.f. Fig. 1(a)). Figure 3a shows a representative electron diffraction pattern of the $WSe_2$ flake and the distinctive Bragg diffraction spots that are detected. The diffraction intensity spectrum is obtained by azimuthally averaging the diffraction pattern. The analysis of individual spots yields curves consistent with the averaged data, within the achieved S/N. A typical curve fitting is shown in Figure 3b, from which a one-dimensional diffraction curve is obtained as a function of the scattering vector. The time-dependent changes in the diffraction pattern are then recovered by fitting the location and area of the diffraction peaks at various pump-probe time delays using a piecewise-linear background and a Lorentzian function. The plots of the time-resolved intensity change of the Bragg diffraction spots are shown in Figure 4a at laser excitation fluences of 5.0, 5.9, 7.1, and 8.2 $mJ/cm^2$, respectively.

These temporal profiles are obtained by averaging over the six first-order diffraction spots shown in Figure 3a. The dynamics are well fit using a single exponential function, to obtain a fluence-dependent lifetime attributed to equilibration through electron-phonon and phonon-phonon scattering that results in a coherent response of the lattice following impulsive photo-excitation.

The room temperature thermal conductivity of layered WSe$_2$ is known to be extremely low due to localized lattice vibrations. (46) Hence, the ultrathin monolayer can be heated uniformly using a laser pulse since the spot size of the photo-excitation is much larger than the flake. For these reasons, a two-temperature model can be employed to calculate the rise in temperature of the sample following 100 fs full-width at half maximum (FWHM) incident laser pulse. (47)

Specifically, we numerically solve the relevant two-coupled nonlinear differential equations given by:

$$C_e \frac{\partial}{\partial t} T_e = \frac{\partial}{\partial z}\left(k_e \frac{\partial}{\partial z} T_e\right) - g(T_e - T_l) + S(z,t) \quad (1)$$

$$C_l \frac{\partial}{\partial t} T_l = g(T_e - T_l) \quad (2)$$

where $e$ and $l$ subscripts denote electron and lattice, respectively. $C$ and $k$ are the heat capacities and the thermal conductivity, respectively. $S(z,t)$ is the time- and space-dependent heating term introduced by the femtosecond laser pulse and $g$ is the electron-phonon coupling constant. The values of $k_e, C_l,$ and $C_e$ used are provided in Refs. (48-50). The best fit two-temperature model results are obtained for $g = 1.5 \times 10^6 \, W.m^{-2}.K^{-1}$. In contrast, the electron-phonon coupling constant of a graphene/WSe$_2$ heterostructure is $0.5 \times 10^6 \, W.m^{-2}.K^{-1}$. (51) Here, the principle thermal coupling is to the electron diffraction grid, and the simulations here are meant principally to provide approximate estimates of the thermal evolution of the flake after laser excitation. Figure 5 shows the rise in the electron temperature obtained from the numerical simulation following optical excitation with 5.0 mJ/cm² laser pulses.

The value of the lattice temperature $T_l$ obtained from the simulation is then used to calculate the change in intensity of the Bragg diffraction spots using the Debye-Waller model given by: (52)

$$\frac{I(t)}{I_0} = \exp\left[2s_{hkl}^2 \left(\frac{<u^2(T_0)> - <u^2(T_l)>}{4}\right)\right] \quad (3)$$

$u^2(T)$ is the atomic mean square displacement given by:

$$<u^2(T)> \geq \frac{3\hbar^2}{2mk_b\theta_D}\left[1 + 4\left(\frac{T}{\theta_D}\right)^2 \int_0^{\frac{\theta_D}{T}} \frac{x}{\exp(x)-1} dx\right] \quad (4)$$

where $m$ is the effective mass of the unit cell, $k_b$ is the Boltzmann's constant, and $\theta_D$ is the Debye temperature. As noted above, the simulated change in the intensity of the Bragg diffraction spots show good agreement with the experimental data for $g = 1.5 \times 10^6 \, W.m^{-2}.K^{-1}$. Figures 6a and 6b show the experimental and calculated change in the Bragg diffraction spot intensities derived from eq. (3).

Figure 4b shows a plot of the lifetimes obtained from the single-exponential fit of the transient intensity change versus laser excitation fluence, along with corresponding peak lattice temperatures. These lifetimes can be described using a single relaxation time approximation of the Boltzmann transport equation (BTE) that neglects phonon reabsorption due to their low probability of occurrence. (53-55) The average relaxation lifetime of the phonon modes in 2D materials is given by: (34)

$$\tau \propto \frac{1}{1 + \alpha T} \quad (5)$$

where $\alpha = \frac{2K_b}{\hbar\omega}$, $K_b$ is Boltzmann constant, and $\hbar\omega$ is LO-phonon energy. The calculated lifetime values shown in Figure 4b are calculated for $\hbar\omega = 31.25$ meV, which corresponds to the LO-phonon energy in monolayer WSe$_2$. (7,56) The calculated results from this scaling equation use the known behavior at room temperature, are in reasonable agreement with the experimental lifetime values at the various temperatures predicted from the simple models outlined above.

The time-resolved expansion shown in Figure 7a of the unit cell is calculated from the change of the distances of the three diagonals between the diffraction spots in the hexagonal lattice at different pump-probe time delays. The lattice expansion reaches a maximum at around 20 ps, followed by a contraction back to the relaxed state after 50 ps. The increase of the distance of the diagonal formed between the (-1010) and (10-10) diffraction spots reaches a maximum amplitude of $(18.2 \pm 0.5) \times 10^{-3}$ Å, while the increase of the distances of the two diagonals formed between the (-1100) and (1-100) versus the (01-10) and (0-110) diffraction spots reaches maximum amplitudes of $(15.1 \pm 0.3) \times 10^{-3}$ Å and $(14.9 \pm 0.3) \times 10^{-3}$ Å, respectively. This anisotropy, or non-uniform expansion of the unit cell, breaks the symmetry of the hexagonal structure at ultrafast time scales following laser excitation. For reference, the lattice constant of WSe$_2$ is 3.280 Å.

At the electron energies used here, the radius of the Ewald sphere is very large compared to the reciprocal lattice. Hence, the circumference of the sphere is nearly flat in the transmission mode of the UED instrument. Any tilts of the sample, either static or dynamic, thus act to broaden the diffraction peaks, not move them. (57) Especially when considering normalized distances, as we do here, the dynamic anisotropic expansion behavior observed, and that of the diffraction spot intensities, is inconsistent with sample tilt. We discuss these results and their possible interpretation more fully below.

## Discussion

Estimates of the thermal expansion coefficient of WSe$_2$ can be obtained from the quasi-harmonic approximation using first-principle calculations and the linear thermal expansion coefficient, $a_i(T)$, given by: (58)

$$a_i(T) = \left(1 + \alpha_{ij}(T)T\right)a_j(T_0) \quad (6)$$

where $\alpha_{ij}(T)$ is the linear thermal expansion tensor and $a_j(T_0)$ is the j component of lattice vector at reference temperature $T_0$. The calculated mean value of the thermal expansion coefficient is $<a(T)> = 0.5501 \times 10^{-5} K^{-1}$. (58) From the thermal expansion coefficient the *average* expansion of the WSe$_2$ unit cell at various temperatures above 200 K can be predicted. Figure 7b shows the plot of the average expansion versus fluence and calculated temperature values, and the good agreement that results.

The thermal expansion of the unit cell following photo-excitation is known to have distinct effects on the band structure in WSe$_2$. (40) In particular, the highly anisotropic lattice expansion is the driving force behind the K → Γ transition in the band structure, which occurs as a result of the indirect optical transition found in few-layer or monolayer WSe$_2$. The large anisotropy in the lattice deformation leads to a clear breaking of inversion symmetry of the unit cell. Ultrafast diffuse electron scattering measurements in bulk WSe$_2$ illustrate the key role of electron-phonon coupling and intervalley scattering, with relaxation of the initially hot photoelectrons occurring over a timescale of 10's of picoseconds, (59) similar to that observed here in monolayer WSe$_2$.

More generally, the breaking of inversion symmetry in monolayer TMDCs can give rise to an optically-induced valley Hall effect which forms the basis of novel spintronic applications. (60-61) Our findings shed light on the nature and time scale of transient structural changes within the unit cell that accompany this band structure evolution following laser excitation in WSe2, which significantly impacts the electrical and optical properties of atomically-thin semiconductor TMDCs and monolayers.

Interestingly, the lattice anisotropy evolution is found to be ~commensurate with the thermal evolution of the flake. Here, static measurements on strain-engineered flake geometries may shed light on the observed ultrafast behavior. In such experiments, the reported degenerate, thickness-dependent frequency trends in the Raman active in-plane $E_{2g}^1$ mode may are likely critical. (62,63) The application of roughly 1-3% uniaxial strain is found to break this degeneracy, and leads to a peak splitting into $E_{2g}^{1+}$ and $E_{2g}^{1-}$ components, to higher and lower frequencies (64) – with the $E_{2g}^{1-}$ peak shifting by a larger amount than does the $E_{2g}^{1+}$ peak. The symmetric A$_{1g}$ mode also becomes visible in polarized Raman measurements. Both trends are illustrative of a deformation of the unit cell to lower symmetry, while the tensile and compressive shifts observed are indicative of an increase and reduction of the lattice constant, respectively. Moreover, Raman shifts are indicative of the amount by which the lattice constant changes. Hence, any applied axial strain on monolayer WSe2 will lead to an overall increase in the lattice constant since the tensile shift to lower frequencies is larger than the compressive shift to higher frequencies. (62) And, since these shifts are directional, the increase in the lattice constant is expected to be anisotropic. Depending on the mounting approach, monolayer flakes of WSe2 are often found to be under small amounts of axial strain, even in vacuum. Here, we do see small red shifts of the photoluminescence spectrum that are indicative of low levels of strain (65) induced by the TEM grid (<0.5%). These level of strain cannot be detected with the UED apparatus, all measurements taken before the 800 nm pump pulse arrives yield symmetric geometries. Upon photo-excitation, however, the focused laser pulse can drive transient charge carrier dynamics and also induce further uniaxial strain of the monolayer. Our results are agnostic as to the mechanism that generates this strain; that is whether ultrafast electron-phonon coupling initiates deformation or whether macroscopic forces induced by the mounting of the flake onto the TEM grid lead to the breaking of the degeneracy of the in-plane phonon modes. The available data only suggests at present an ultrafast anisotropic lattice expansion on top of the general expansion generated by the heating of the sample, and we measure the duration of this effect, which impacts the nature of optically-induced transient behavior in the monolayer flake. Because phonon-phonon coupling is important to both processes, the observed anisotropic expansion persists over similar lifetimes to that of the electron-phonon and phonon-phonon scattering probed by the change in the intensity of the Bragg diffraction spots. Further highly time resolved measurements of the initial rise in the lattice size would be needed to address strain versus intrinsic material processes in the observed anisotropic expansion of monolayer $WSe_2$.

## Conclusions

In summary, ultrafast electron diffraction is used to investigate the transient structural dynamics in a monolayer flake of $WSe_2$ with both high spatial and temporal resolution. The change in the intensity of the Bragg diffraction spots follows single-exponential dynamics that is attributed to electron-phonon and phonon-phonon equilibration, whose temporal behavior is stable over the long illumination times of the experiment. Similar to the behavior observed in graphene, the experimental lifetimes obtained at various excitation fluences are accurately described by a single-mode relaxation time approximation. And, a long predicted and highly anisotropic ultrafast expansion of the unit cell in semiconductor TMDC monolayers is observed for the first time from the time-resolved lattice expansion. These novel structural changes occur on the picosecond time scale, and lead to the evolution of the band structure that results from the indirect optical transition. Future experiments that manipulate the anisotropic lattice expansion with light, or by applying strain to these 2D materials or changing their composition through chemical doping or by embedding quantum emitters in the monolayers, would provide additional control over the optical transition as well as the mechanical and electronic properties of these samples, and lead to novel applications in optoelectronics and spintronics.

AUTHOR INFORMATION

**Corresponding Author**

*Email: gab@caltech.edu. Phone: (626) 395-6296

**Author Contributions**

The manuscript was written through contributions of all authors. All authors have given approval to the final version of the manuscript.

**Notes**

The authors declare no competing financial interest.

ACKNOWLEDGMENT

Generous financial support for this work was provided by the Gordon and Betty Moore Foundation, as well as the National Science Foundation and the Air Force Office of Scientific Research (grant FA9550-16-1-0200). The authors dedicate this manuscript to Dr. Michael Berman, for his unstinting support of ultrafast electron microscopy through the AFOSR Molecular Dynamics and Theoretical Chemistry program, and to the memory of our colleague and mentor Professor Ahmed H. Zewail.

**Figures and Captions**

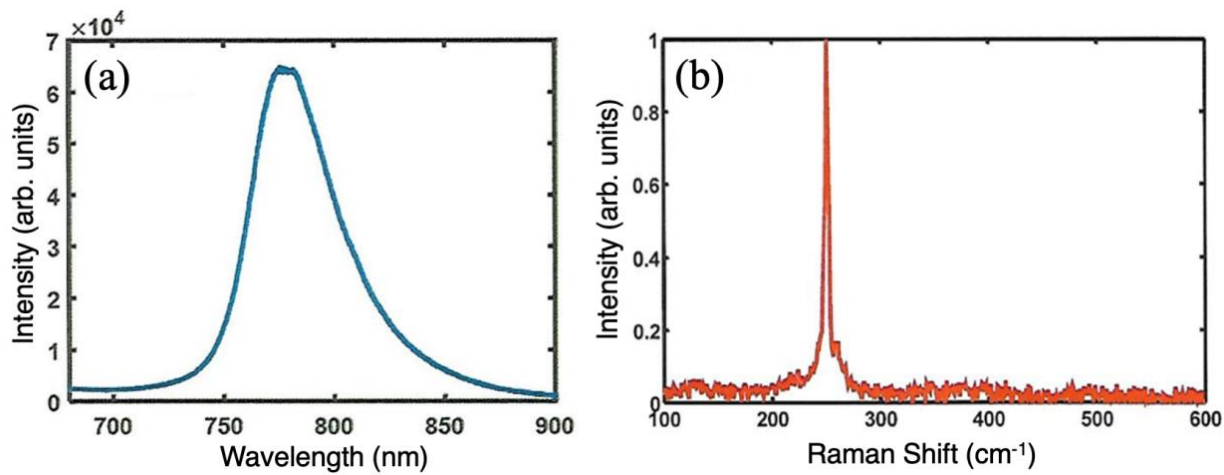

**Figure 1.** (a) Photoluminescence spectrum of a WSe$_2$ monolayer on a sapphire substrate following 532 nm laser excitation (0.1 mW, average power). (b) Raman spectrum of the same WSe$_2$ monolayer, on a sapphire substrate. Excitation at 514 nm.

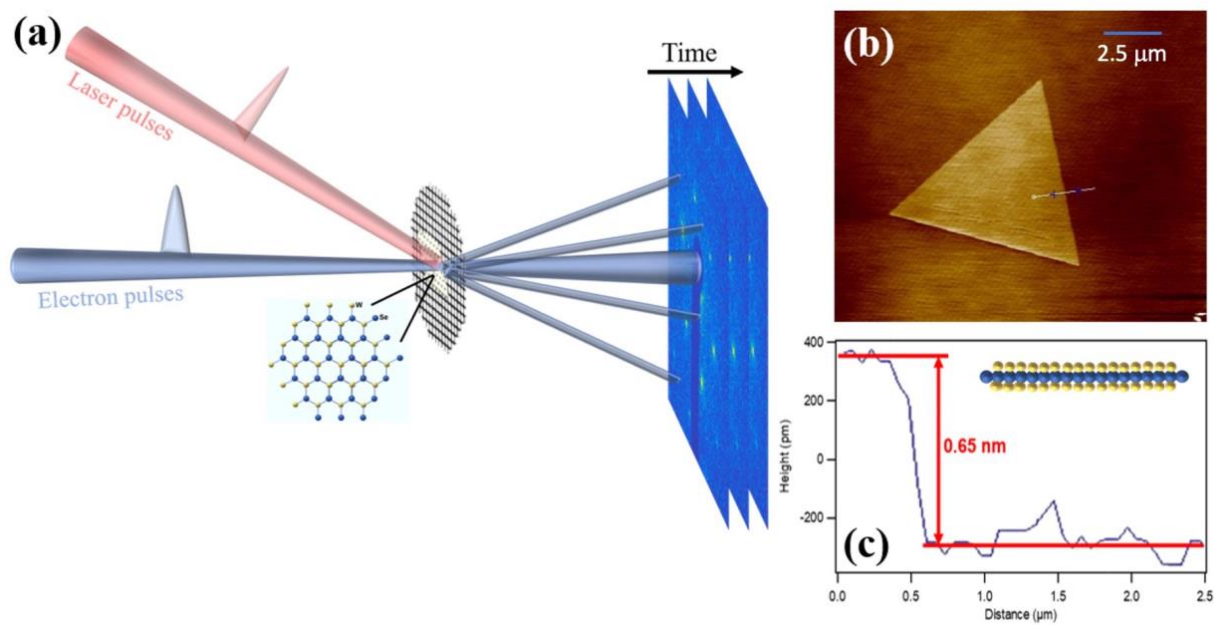

**Figure 2.** (a) Schematic of the UED setup in the transmission geometry showing a spatial overlap between the laser and electron pulses. (b) AFM image of a WSe$_2$ monolayer flake mounted on a sapphire substrate. (d) The height of the sample is 0.65 nm, corresponding to the thickness of a single unit cell.

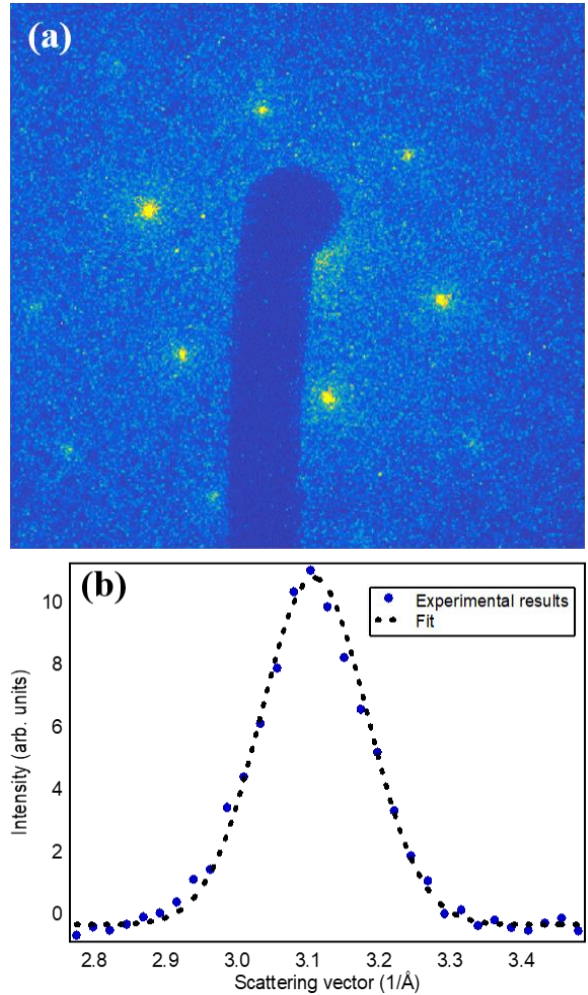

**Figure 3.** (a) Electron diffraction image of the WSe$_2$ monolayer. (b) Diffraction intensity spectrum obtained by azimuthally averaging the diffraction pattern, and its curve fitting.

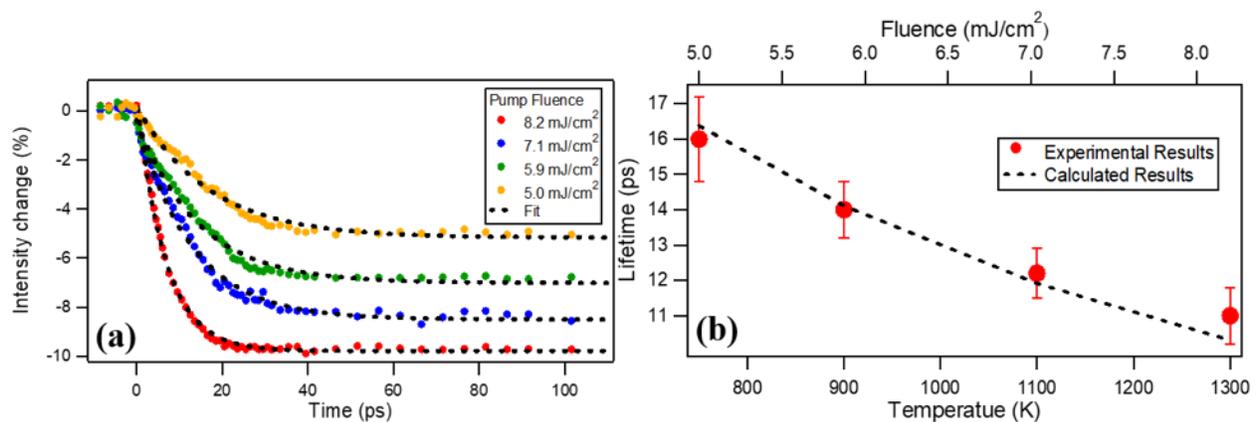

**Figure 4.** (a) Time-resolved diffraction spot intensity change at various laser incident fluences. The dynamics are characterized by a single-exponential fit. (b) Plot of the lifetimes obtained from the single-exponential fit of the transient intensity change versus laser fluence, and the corresponding temperatures estimated from a solution of the 1-D heat diffusion equation.

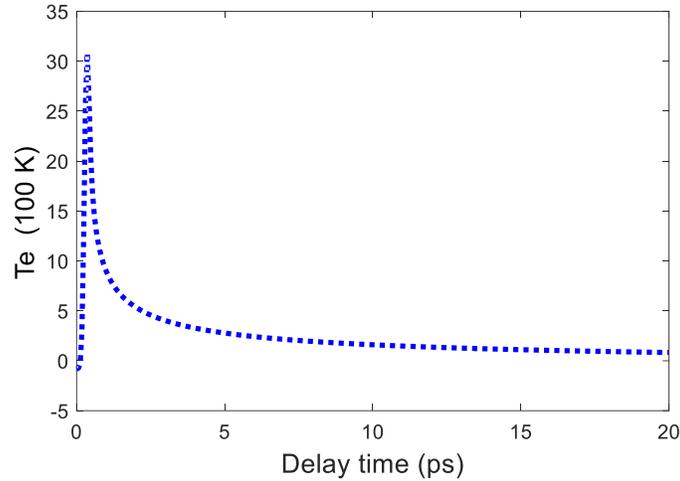

**Figure 5.** The time-dependent rise in the electron temperature obtained from the two-temperature model following excitation by a 5.0 mJ/cm$^2$ laser pulse.

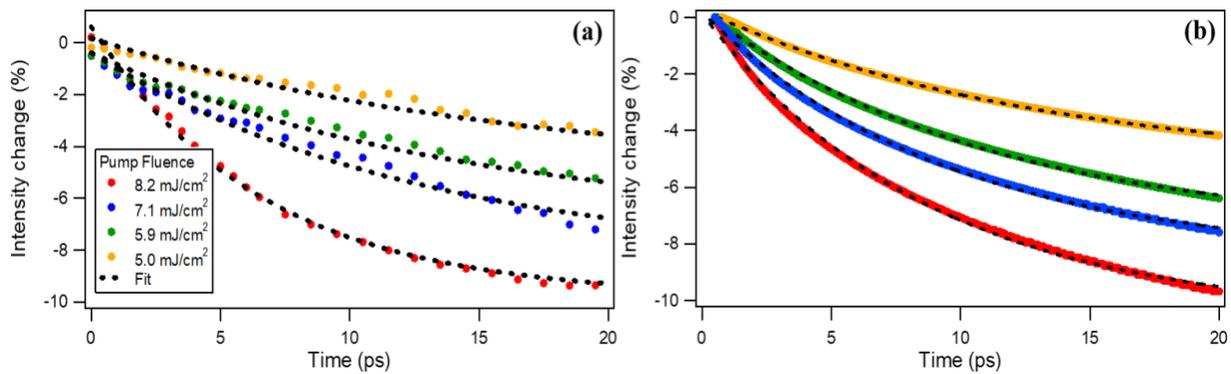

**Figure 6.** (a) Experimental and (b) calculated time-dependent change in intensity of the Bragg diffraction peak of WSe$_2$ monolayer, following eq. (3).

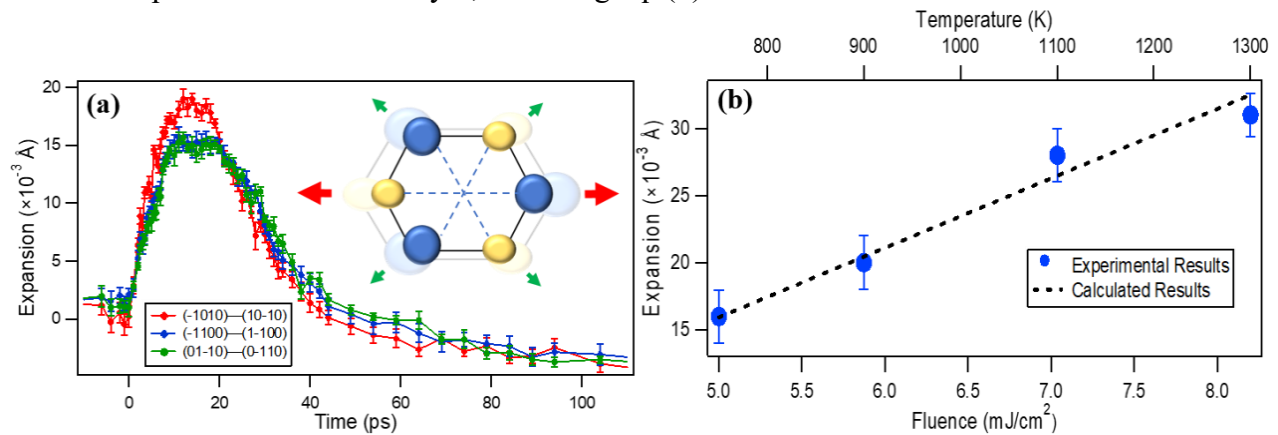

**Figure 7.** (a) Time-resolved anisotropic expansion of the WSe$_2$ unit cell obtained by calculating the change in distance of the diagonals between the Bragg diffraction spots. (b) Plot of the average expansion versus laser fluence and temperatures. The experimental results are overlapped with the calculated expansion values using the quasi-harmonic approximation.

**TOC Graphic**

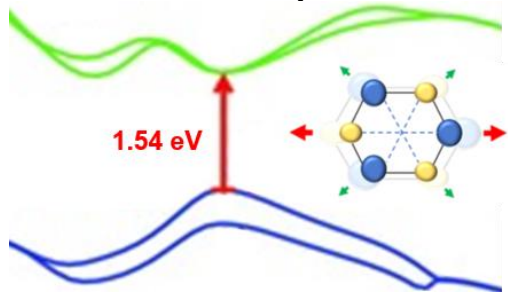